\documentclass[a4paper,11pt]{article}
\usepackage{pos}
\RequirePackage{ifpdf}
\usepackage{graphicx}
\usepackage{amsmath}
\usepackage{xspace}
\usepackage{float}
\usepackage{placeins}
\usepackage{pdflscape}
\usepackage{ctable}
\usepackage{rotating}
\usepackage{multirow}
\usepackage{tabulary}
\usepackage{array,booktabs}
\usepackage{tabularx}
\newcolumntype{Y}{>{\centering\arraybackslash}X}

\usepackage{cleveref}

\crefname{table}{Tab.}{Tab.}

\title{ALICE data processing for Run 3 and Run 4 at the LHC}

\author*[a]{Chiara Zampolli for the ALICE Collaboration}

\affiliation[a]{CERN,\\
  Espl. des Particules 1, 1211 Meyrin, Switzerland}

\emailAdd{Chiara.Zampolli@cern.ch}

\abstract{During the upcoming \run{3} and \run{4} of the LHC, ALICE will take data at a peak \PbPb collision rate of 50 kHz. This will be made possible thanks to the upgrade of the main tracking detectors of the experiment, and with a new data processing strategy. In order to collect the statistics needed for the precise measurements that ALICE aims at, a continuous readout will be adopted. This brings about the challenge of handling unprecedented data rates. The ~3.5 \TBs of raw data from the detectors will be reduced to about 600 \GBs on the First Level Processing (FLP) nodes, and sent to the Event Processing layer for further processing and reduction to $\sim$100 \GBs of data to be stored permanently. This synchronous processing stage, which will include reconstruction, calibration and compression procedures, will be followed by an asynchronous one to account for final calibrations. Quality Control (QC) will be intensively used in all the processing stages. This contribution illustrates the processing flow for ALICE in \run{3} and \run{4}, with emphasis on the components of the synchronous processing. The chosen software design will be described. An overview of the data analysis framework is included as well.}

\FullConference{%
  40th International Conference on High Energy physics - ICHEP2020\\
  July 28 - August 6, 2020\\
  Prague, Czech Republic (virtual meeting)
}

\begin{document}
\newcommand {\red}[1]    {\textcolor{red}{#1}}
\newcommand {\green}[1]  {\textcolor{green}{#1}}
\newcommand {\blue}[1]   {\textcolor{blue}{#1}}
\newcommand {\dgreen}[1]    {\textcolor{darkgreen}{#1}}
\newcommand {\orange}[1]   {\textcolor{orange}{#1}}

\newcommand{\deutsch}[1]{\foreignlanguage{german}{#1}}

\newcommand{\ToDo}      {\textcolor{blue}{\footnotesize \textsc{ToDo}}}
\newcommand{\TBC}       {\textcolor{red}{\footnotesize \textsc{TBC}}}

\DeclareRobustCommand{\unit}[2][]{%
        \begingroup%
                \def\0{#1}%
                \expandafter%
        \endgroup%
        \ifx\0\@empty%
                \ensuremath{\mathrm{#2}}%
        \else%
                \ensuremath{#1\,\mathrm{#2}}%
        \fi%
        }
\DeclareRobustCommand{\unitfrac}[3][]{%
        \begingroup%
                \def\0{#1}%
                \expandafter%
        \endgroup%
        \ifx\0\@empty%
                \raisebox{0.98ex}{\ensuremath{\mathrm{\scriptstyle#2}}}%
                \nobreak\hspace{-0.15em}\ensuremath{/}\nobreak\hspace{-0.12em}%
                \raisebox{-0.58ex}{\ensuremath{\mathrm{\scriptstyle#3}}}%
        \else
                \ensuremath{#1}\,%
                \raisebox{0.98ex}{\ensuremath{\mathrm{\scriptstyle#2}}}%
                \nobreak\hspace{-0.15em}\ensuremath{/}\nobreak\hspace{-0.12em}%
                \raisebox{-0.58ex}{\ensuremath{\mathrm{\scriptstyle#3}}}%
        \fi%
}

%
%
\newcommand{\ie}{i.\,e.\;}
\newcommand{\eg}{e.\,g.\;}

\newcommand{\run}[1]{\textsc{Run\,#1}}
\newcommand{\oo}{\mbox{${\rm O}^2$}\xspace}

%
%

\newcommand{\dNdeta}{\ensuremath{\mathrm{d}N_{\rm ch}/\mathrm{d}\eta}\xspace}

%
%
\newlength{\smallerpicsize}
\setlength{\smallerpicsize}{70mm}
\newlength{\smallpicsize}
\setlength{\smallpicsize}{90mm}
\newlength{\mediumpicsize}
\setlength{\mediumpicsize}{120mm}
\newlength{\largepicsize}
\setlength{\largepicsize}{150mm}

\newcommand{\PICX}[5]{
   \begin{figure}[!hbt]
      \begin{center}
         \vspace{3ex}
         \includegraphics[width=#3]{#1}
         \caption[#4]{\label{#2} #5}        
      \end{center}  
   \end{figure}
}

\newcommand{\PICH}[5]{
   \begin{figure}[H]
      \begin{center}
         \vspace{3ex}
         \includegraphics[width=#3]{#1}
         \caption[#4]{\label{#2} #5}        
      \end{center}  
   \end{figure}
}

%
%
%
\newcommand{\figs}{Figs.\xspace}
\newcommand{\Figs}{Figures\xspace}
\newcommand{\eqn}{equation\xspace}
\newcommand{\Eqn}{Equation\xspace}
\newcommand{\figref}[1]{Fig.~\ref{#1}}
\newcommand{\Figref}[1]{Figure~\ref{#1}}
\newcommand{\tabref}[1]{Tab.~\ref{#1}}
\newcommand{\Tabref}[1]{Table~\ref{#1}}
\newcommand{\appref}[1]{appendix~\ref{#1}}
\newcommand{\Appref}[1]{Appendix~\ref{#1}}
\newcommand{\secs}{Secs.\xspace}
\newcommand{\Secs}{Sections\xspace}
\newcommand{\secref}[1]{Sec.~\ref{#1}}
\newcommand{\Secref}[1]{Section~\ref{#1}}
\newcommand{\chaps}{Chaps.\xspace}
\newcommand{\Chaps}{Chapters\xspace}
\newcommand{\chapref}[1]{Chap.~\ref{#1}}
\newcommand{\Chapref}[1]{Chapter~\ref{#1}}
\newcommand{\lstref}[1]{Listing~\ref{#1}}
\newcommand{\Lstref}[1]{Listing~\ref{#1}}
%
%
\newcommand{\otoprule}{\midrule[\heavyrulewidth]}
\topfigrule
%
\newcommand {\stat}     {({\it stat.})~}
\newcommand {\syst}     {({\it syst.})~}
 \newcommand {\mom}       {\ensuremath{p}}
\newcommand {\pT}        {\pt}
\newcommand {\meanpT}    {\ensuremath{\langle p_{\mathrm{T}} \kern-0.1em\rangle}\xspace}
\newcommand {\mean}[1]   {\ensuremath{\langle #1 \kern-0.1em\rangle}\xspace} 
\newcommand {\sqrtsNN}   {\ensuremath{\sqrt{s_{\textsc{NN}}}}\xspace}
\newcommand {\sqrts}     {\ensuremath{\sqrt{s}}\xspace}
\newcommand {\vf}        {\ensuremath{v_{\mathrm{2}}}\xspace}
\newcommand {\et}        {\ensuremath{E_{\mathrm{t}}}\xspace}
\newcommand {\mT}        {\ensuremath{m_{\mathrm{T}}}\xspace}
\newcommand {\mTmZero}   {\ensuremath{m_{\mathrm{T}} - m_0}\xspace}
\newcommand {\minv}      {\mbox{$m_{\ee}$}}
\newcommand {\rap}       {\mbox{$y$}}
\newcommand {\absrap}    {\mbox{$\left | y \right | $}}
\newcommand {\rapXi}     {\mbox{$\left | y(\rmXi) \right | $}}
\newcommand {\abspseudorap} {\mbox{$\left | \eta \right | $}}
\newcommand {\pseudorap} {\mbox{$\eta$}}
\newcommand {\cTau}      {\ensuremath{c\tau}}
\newcommand {\sigee}     {$\sigma_E$/$E$}
\newcommand {\dNdy}      {\ensuremath{\mathrm{d}N/\mathrm{d}y}}
\newcommand {\dNdpt}     {\ensuremath{\mathrm{d}N/\mathrm{d}\pT }}
\newcommand {\dNdptdy}   {\ensuremath{\mathrm{d^{2}}N/\mathrm{d}\pT\mathrm{d}y }}
\newcommand {\fracdNdptdy}   {\ensuremath{ \frac{\mathrm{d^{2}}N}{\mathrm{d}\pT\mathrm{d}y } }}
\newcommand {\dNdmtdy}   {\ensuremath{\mathrm{d^{2}}N/\mathrm{d}\mT\mathrm{d}y }}
\newcommand {\dN}        {\ensuremath{\mathrm{d}N }}
\newcommand {\dNsquared} {\ensuremath{\mathrm{d^{2}}N }}
\newcommand {\dpt}       {\ensuremath{\mathrm{d}\pT }}
\newcommand {\dy}        {\ensuremath{\mathrm{d}y}}
\newcommand {\dNdyBold}  {\ensuremath{\boldsymbol{\dN/\dy}}\xspace}
\newcommand {\dNchdy}    {\ensuremath{\mathrm{d}N_\mathrm{ch}/\mathrm{d}y }\xspace}
\newcommand {\dNchdeta}  {\ensuremath{\mathrm{d}N_\mathrm{ch}/\mathrm{d}\eta }\xspace}
\newcommand {\dNchdptdeta}  {\ensuremath{\mathrm{d}N_\mathrm{ch}/\mathrm{d}\pT\mathrm{d}\eta }\xspace}
\newcommand {\Raa}       {\ensuremath{R_\mathrm{AA}}}
\newcommand {\RpPb}       {\ensuremath{R_\mathrm{pPb}}\xspace}
\newcommand {\Nevt}      {\ensuremath{N_\mathrm{evt}}}
\newcommand {\NevtINEL}  {\ensuremath{N_\mathrm{evt}(\textsc{inel})}}
\newcommand {\NevtNSD}   {\ensuremath{N_\mathrm{evt}(\textsc{nsd})}}
\newcommand{\dEdx}       {\ensuremath{\mathrm{d}E/\mathrm{d}x}\xspace}
\newcommand{\ttof}       {\ensuremath{t_\mathrm{TOF}}\xspace}
\newcommand {\ee}        {\mbox{$\mathrm {e^+e^-}$}\xspace}
\newcommand {\ep}        {\mbox{$\mathrm {e\kern-0.05em p}$}\xspace}
\newcommand {\pp}        {\mbox{$\mathrm {p\kern-0.05em p}$}\xspace}
\newcommand {\ppBoldMath} {\mbox{$\mathrm { \mathbf p\kern-0.05em \mathbf p }$}\xspace}
\newcommand {\ppbar}     {\mbox{$\mathrm {p\overline{p}}$}\xspace}
\newcommand {\PbPb}      {\ensuremath{\mbox{Pb--Pb}}\xspace}
\newcommand {\AuAu}      {\ensuremath{\mbox{Au--Au}}\xspace}
\newcommand {\CuCu}      {\ensuremath{\mbox{Cu--Cu}}\xspace}
\renewcommand {\AA}      {\ensuremath{\mbox{A--A}}\xspace}
\newcommand {\pA}        {\ensuremath{\mbox{p--A}}\xspace}
\newcommand {\pPb}       {\ensuremath{\mbox{p--Pb}}\xspace}
\newcommand {\Pbp}       {\ensuremath{\mbox{Pb--p}}\xspace}
\newcommand {\hPM}       {\ensuremath{h^{\pm}}\xspace}
\newcommand {\rphi}      {\ensuremath{(r,\phi)}\xspace}
\newcommand {\alphaS}    {\ensuremath{ \alpha_s}\xspace}
\newcommand {\MeanNpart} {\mbox{\ensuremath{< \kern-0.15em N_{part} \kern-0.15em >}}}

\newcommand {\sig}       {\ensuremath{S}\xspace}
\newcommand {\expsig}    {\ensuremath{\hat{S}}\xspace}
\newcommand {\prob}      {\ensuremath{P}\xspace}
\newcommand {\prior}     {\ensuremath{C}\xspace}
\newcommand {\prop}      {\ensuremath{F}\xspace}
\newcommand {\atrue}     {\ensuremath{\vec{A}_{\mathrm{true}}}\xspace}
\newcommand {\ameas}     {\ensuremath{\vec{A}_{\mathrm{meas}}}\xspace}
\newcommand {\detresp}   {\ensuremath{R}\xspace}

\newcommand {\pid}       {\ensuremath{\mathrm{\epsilon}_\mathrm{PID}}\xspace}
\newcommand {\nsigma}    {\ensuremath{\mathrm{n_{\sigma}}}\xspace}
\newcommand {\ylab} {\ensuremath{\mathrm{| y_{lab} |}}\xspace}

%
%
\newcommand {\mass}     {\mbox{\rm MeV$\kern-0.15em /\kern-0.12em c^2$}}
\newcommand {\tev}      {\mbox{${\rm TeV}$}\xspace}
\newcommand {\gev}      {\mbox{${\rm GeV}$}\xspace}
\newcommand {\mev}      {\mbox{${\rm MeV}$}\xspace}
\newcommand {\kev}      {\mbox{${\rm keV}$}\xspace}
\newcommand {\tevBoldMath}  {\mbox{${\rm \mathbf{TeV}}$}}
\newcommand {\gevBoldMath}  {\mbox{${\rm \mathbf{GeV}}$}}
\newcommand {\mmom}     {\mbox{\rm MeV$\kern-0.15em /\kern-0.12em c$}}
\newcommand {\gmom}     {\mbox{\rm GeV$\kern-0.15em /\kern-0.12em c$}}
\newcommand {\mmass}    {\mbox{\rm MeV$\kern-0.15em /\kern-0.12em c^2$}}
\newcommand {\gmass}    {\mbox{\rm GeV$\kern-0.15em /\kern-0.12em c^2$}}
\newcommand {\nb}       {\mbox{\rm nb}}
\newcommand {\musec}    {\mbox{$\mu {\rm s}$}}
\newcommand {\nsec}     {\mbox{${\rm ns}$}}
\newcommand {\psec}     {\mbox{${\rm ps}$}}
\newcommand {\fmC}      {\mbox{${\rm fm/c}$}}
\newcommand {\fm}       {\mbox{${\rm fm}$}}
\newcommand {\cm}       {\mbox{${\rm cm}$}}
\newcommand {\mm}       {\mbox{${\rm mm}$}}
\newcommand {\mim}      {\mbox{$ \mu {\rm m}$}}
\newcommand {\cmq}      {\mbox{${\rm cm}^{2}$}}
\newcommand {\mmq}      {\mbox{${\rm mm}^{2}$}}
\newcommand {\dens}     {\mbox{${\rm g}/{\rm cm}^{3}$}}
\newcommand {\lum}      {\, \mbox{${\rm cm}^{-2} {\rm s}^{-1}$}}
\newcommand {\barn}     {\, \mbox{${\rm barn}$}}
\newcommand {\m}        {\, \mbox{${\rm m}$}}
\newcommand {\dg}       {\mbox{$\kern+0.1em ^\circ$}}
\newcommand{\mpp}{\ensuremath{\mathrm{pp}}\xspace}
\newcommand{\rts}{\ensuremath{\sqrt{s}}\xspace}
\newcommand{\GeV}{\ensuremath{\mathrm{GeV}}\xspace}
\newcommand{\TeV}{\ensuremath{\mathrm{TeV}}\xspace}
\newcommand{\gevc}{GeV/\ensuremath{c}\xspace}
\newcommand{\GeVc}{\gevc}
\newcommand{\mevc}{\ensuremath{\mathrm{MeV}/c}\xspace}
\newcommand{\mevcc}{\ensuremath{\mathrm{MeV}/c^{2}}\xspace}
\newcommand{\gevcc}{\ensuremath{\mathrm{GeV}/c^{2}}\xspace}
\newcommand{\pt}{\ensuremath{p_{\rm T}}\xspace}
\newcommand{\kt}{\ensuremath{k_{\rm T}}\xspace}
\newcommand {\lumi}{\mathcal{L}_{\rm int}\xspace}
\newcommand{\nbinv}{\ensuremath{\rm nb^{-1}}}
\newcommand {\ubinv}{\ensuremath{\mu\rm b^{-1}}}
\newcommand {\um}{\ensuremath{\mu\rm m}\xspace}
\newcommand {\GBs}      {\mbox{${\rm GB/s}$}\xspace}
\newcommand {\TBs}      {\mbox{${\rm TB/s}$}\xspace}

\newcommand{\lt}{\textless}
\newcommand{\ctau}{\ensuremath{c\tau\xspace}}

%
%

\newcommand{\ePlusMinus}       {\mbox{$\mathrm {e^{\pm}}$}\xspace}
\newcommand{\muPlusMinus}      {\mbox{$\mathrm {\mu^{\pm}}$}\xspace}

\newcommand{\pion}            {\mbox{$\mathrm {\pi}$}\xspace}
\newcommand{\piZero}            {\mbox{$\mathrm {\pi^0}$}\xspace}
\newcommand{\piMinus}           {\ensuremath{\mathrm {\pi^-}}\xspace}
\newcommand{\piPlus}            {\ensuremath{\mathrm {\pi^+}}\xspace}
\newcommand{\piPlusMinus}       {\mbox{$\mathrm {\pi^{\pm}}$}\xspace}

\newcommand{\proton}    {\mbox{$\mathrm {p}$}\xspace}
\newcommand{\pbar}      {\mbox{$\mathrm {\overline{p}}$}\xspace}
\newcommand{\pOuPbar}   {\mbox{$\mathrm {p^{\pm}}$}\xspace}
\newcommand{\DZero}     {\mbox{$\mathrm {D^0}$}\xspace}
\newcommand{\DZerobar}  {\mbox{$\mathrm {\overline{D}^0}$}\xspace}
\newcommand{\Bminus}    {\mbox{$\mathrm {B^-}$}\xspace}
\newcommand{\BZero}     {\mbox{$\mathrm {B^0}$}\xspace}
\newcommand{\BZerobar}  {\mbox{$\mathrm {\overline{B}^0}$}\xspace}

\newcommand{\Dmes}       {\mbox{$\mathrm {D}$}\xspace}
\newcommand{\Lc}         {\mbox{$\mathrm {\Lambda_{c}}$}\xspace}
\newcommand{\Lb}{\ensuremath{\rm {\Lambda_b}}\xspace}
\newcommand{\Xic}         {\mbox{$\mathrm {\Xi_{c}}$}\xspace}
\newcommand{\lambdab}     {\mbox{$\mathrm {\Lambda_{b}^{0}}$}\xspace}
\newcommand{\lambdac}     {\mbox{$\mathrm {\Lambda_{c}^{+}}$}\xspace}
\newcommand{\xicz}        {\mbox{$\mathrm {\Xi_{c}^{0}}$}\xspace}
\newcommand{\xiczp}        {\mbox{$\mathrm {\Xi_{c}^{0,+}}$}\xspace}
\newcommand{\xicp}        {\mbox{$\mathrm {\Xi_{c}^{+}}$}\xspace}
\newcommand{\xib}        {\mbox{$\mathrm {\Xi_{b}}$}\xspace}
\newcommand{\LambdaParticle}        {\mbox{$\mathrm {\Lambda}$}\xspace}

\newcommand{\rmLambdaZ}         {\mbox{$\mathrm {\Lambda}$}\xspace}
\newcommand{\rmAlambdaZ}        {\mbox{$\mathrm {\overline{\Lambda}}$}\xspace}
\newcommand{\rmLambda}          {\mbox{$\mathrm {\Lambda}$}\xspace}
\newcommand{\rmAlambda}         {\mbox{$\mathrm {\overline{\Lambda}}$}\xspace}
\newcommand{\rmLambdas}         {\mbox{$\mathrm {\Lambda \kern-0.2em + \kern-0.2em \overline{\Lambda}}$}\xspace}

\newcommand{\Vzero}             {\mbox{$\mathrm {V^0}$}\xspace}
\newcommand{\Vzerob}             {\mbox{{\bold $\mathrm {V^0}$}}\xspace}
\newcommand{\Kzero}             {\mbox{$\mathrm {K^0}$}\xspace}
\newcommand{\Kzs}               {\ensuremath{\mathrm {K^0_S}}\xspace}
\newcommand{\phimes}            {\ensuremath{\mathrm {\phi}}\xspace}
\newcommand{\Kminus}            {\ensuremath{\mathrm {K^-}}\xspace}
\newcommand{\Kplus}             {\ensuremath{\mathrm {K^+}}\xspace}
\newcommand{\Kstar}             {\mbox{$\mathrm {K^*}$}\xspace}
\newcommand{\Kplusmin}          {\mbox{$\mathrm {K^{\pm}}$}\xspace}
\newcommand{\Jpsi}              {\ensuremath{\rm J/\psi}\xspace}
\newcommand{\DtoKpi}{\ensuremath{\rm D^0\to K^-\pi^+}\xspace}
\newcommand{\DtoKpipi}{\ensuremath{\rm D^+\to K^-\pi^+\pi^+}\xspace}
\newcommand{\DstartoDpi}{\ensuremath{\rm D^{*+}\to D^0\pi^+}\xspace}
\newcommand{\Dzero}{\ensuremath{\mathrm {D^0}}\xspace}
\newcommand{\Dzerobar}{\ensuremath{\mathrm{\overline{D}^0}}\xspace}
\newcommand{\Dstar}{\ensuremath{\rm D^{*+}}\xspace}
\newcommand{\Dplus}{\ensuremath{\rm D^+}\xspace}
\newcommand{\decleng}{\ensuremath{\rm L}_{xyz}}
\newcommand{\Lcminus}{\ensuremath{\rm {\overline{\Lambda}{}_c^-\xspace}}}
\newcommand{\Lcplus}{\ensuremath{\rm {\Lambda_c^+}\xspace}}
\newcommand{\Lbzero}{\ensuremath{\rm {\Lambda_b^0}}\xspace}
\newcommand{\LctopKpi}{\ensuremath{\rm \Lambda_{c}^{+}\to p K^-\pi^+}\xspace}
\newcommand{\LbtoLc}{\ensuremath{\rm \Lambda_{b}^{0}\to \Lc + \rm{X}}\xspace}
\newcommand{\LctopKzS}{\ensuremath{\rm \Lambda_{c}^{+}\to p K^{0}_{S}}\xspace}
\newcommand{\LctoenuLambda}{\ensuremath{\rm \Lambda_{c}^{+}\to e^{+} \nu_{e} \Lambda}\xspace}
\newcommand{\cosP}{\ensuremath{\rm cos_{\Theta_{pointing}}}\xspace}
\newcommand{\KzStopippim}{\ensuremath{\rm K^{0}_{S}\to \pi^{+} \pi^{-}}\xspace}
\newcommand{\Lambdatoppim}{\ensuremath{\rm \Lambda \to p \pi^{-}}\xspace}
\newcommand{\nue}{$\nu_e$}
\newcommand{\DtopiKzs}{\ensuremath{\rm D^+\to \pi^+ K^{0}_{S}}\xspace}
\newcommand{\DstoKKzs}{\ensuremath{\rm D_s^+\to K^+ K^{0}_{S}}\xspace}

\newcommand{\ptLc}{\ensuremath{p_{\rm T, \Lambda_c}}\xspace}
\newcommand{\ptpion}{\ensuremath{p_{\rm T, \pi}}\xspace}
\newcommand{\ptK}{\ensuremath{p_{\rm T, K}}\xspace}
\newcommand{\ptproton}{\ensuremath{p_{\rm T, \proton}}\xspace}

\maketitle

\section{ALICE upgrade}
\label{sec: ALICE upgrade}
A Large Ion Collider Experiment~\cite{Aamodt:2008zz}, \cite{Abelev:2014ffa} is the experiment at the CERN LHC dedicated to the study of heavy-ion collisions. 
In order to cope with the very large track density of such 
environment, which reaches $\sim$2000 charged particles per unit rapidity in the most central collisions~\cite{Adam:2015ptt}, ALICE is endowed with optimized and 
precise tracking
detectors, covering the pseudorapidity region $|\eta| < 0.9$ and full azimuth, which allow for track reconstruction down to very low momenta, of the order 
of a few hundreds of \mev.  The main ALICE tracking device in the central region is the TPC, a large Time Projection Chamber based on multi-proportional wire chambers,
recording up to 159 reconstructed points for charged tracks. Thanks to the diverse particle identification capabilities of many of its detectors, 
ALICE identifies hadrons from the very low momenta up to a few \gevc. 
During the very successful \run{2} data taking, ALICE collected $\sim$1~\nbinv~\PbPb events, running at an Interaction Rate (IR) up to $\sim$10 kHz, limited 
by the Time Projection Chamber
readout capability ($< 3.5$ kHz) and the bandwidth of the optical links used by the online data acquisition of the experiment ($\sim$80 GB/s). 
The goal of \run{3} and \run{4} is to collect $\sim$10~\nbinv of \PbPb collisions at the nominal magnetic field of 0.5 T, in addition to $\sim$3~\nbinv at lower B, 
B = 0.2~T. In order to be able to 
record this large volume of data, ALICE will run at a peak interaction rate of 50 kHz, and will undergo very important upgrades involving the data taking paradigm, the detectors, 
and the data processing~\cite{Antonioli:2013ppp}, \cite{Abelevetal:2014cna}, \cite{Buncic:2015ari}. The \run{1} and \run{2} trigger-based data taking will be replaced by a continuous readout mode both for pp and heavy-ion collisions, which imposes drastic changes in some of the detectors. 
The upgrade of the ALICE Inner Tracking System (ITS)~\cite{Abelevetal:2014dna}, the TPC~\cite{ALICE:2014qrd}, \cite{TheALICECollaboration:2015xke} and the new Muon Forward Tracker (MFT)~\cite{MFT} are discussed in the contribution \cite{StefanoPanebianco}. Never the less, here we are going 
to briefly summarize the main changes that the ALICE TPC will go through, since these are the key elements to understand the 
data processing chain of the experiment in \run{3} and \run{4}. 

In order to be operated in continuous data taking mode, the ALICE TPC will replace the multi-proportional wire chambers (having the strong limitation
of applying a gating grid on a trigger signal to prevent ions produced during the amplification process to enter the gas volume) with Gas Electron Multiplier (GEM)
readout chambers. The final configuration of the TPC was defined after intense R\&D studies, and led to GEM stacks made of 4 layers working under  
highly optimized high voltage settings~\cite{Munzer:2020nwl}, which ensure an ion back flow, a tracking performance and energy loss determination as defined in the 
ALICE upgrade requirements~\cite{ALICE:2014qrd}. 

The data processing strategy adopted by ALICE in \run{1} and \run{2} won't fit anymore the data taking conditions of the experiment. The extremely large data size 
that will be read out by the detectors ($\sim$3.5~TB/s) dictates the need to adopt data reduction techniques during the data taking itself, synchronously. 
A total compression factor of 35 will be needed in order to be able to store on a 60 PB buffer disk the data collected during a \PbPb data taking period 
(few weeks per year), achievable 
only if, in addition to data format optimization techniques strongly improved and adapted with 
respect to \run{1} and \run{2} (\eg zero suppression, entropy reduction \ldots ~\cite{MichaelLettrich}) , the output of online reconstruction and calibration will also be used. The ALICE Online-Offline (\oo) processing 
is the synergy of the two online and offline worlds that existed as separate entities in \run{1} and \run{2}, and will work as a whole during \run{3} and \run{4}.

\section{Synchronous versus asynchronous data processing}
\label{sec:Synchronous_versus_asynchronous data processing}
The ALICE \run{3} and \run{4} data processing flow can be divided in two phases: a synchronous processing, occurring during the data taking, and an asynchronous one, happening
with a few weeks delay with respect to the data taking. The ultimate goal of the synchronous processing will be to reduce the size of the 
incoming data from the detectors' readout links such to be able to store them on disk. 

Data will come at a speed of 3.5 TB/s to the First Layer Processing (FLP) nodes, which constitute the first processing level of the \oo farm. Here, zero
suppression will happen inside FPGA-based readout cards (Common Readout Unit, CRU, \cite{Antonioli:2013ppp}, \cite{Mitra:2016tfb}). The data coming 
continuously from the detectors will be chopped on each FLP into 128 $\--$ 256 orbits (corresponding to $\sim$10$ \-- 20$ ms) data packets called 
sub-Time Frames (TF). The name refers to the fact that they will contain the data coming from only a subset of the readout links of the experiment. 
At this stage, the data rate will 
be reduced by a factor of $< 6$, with $\sim$ 0.6 TB/s reaching the second processing level, happening on the Event Processing Nodes (EPN). On the EPNs, the
full TF will be built, merging the data from the different sub-time frames. Synchronous reconstruction, calibration, and data compression will be
performed, resulting in the Compressed Time Frames (CTFs) stored on a 60 PB disk buffer at a data rate of $< 0.1$ TB/s. The total reduction 
during synchronous processing results in a factor of 35. 

The \oo farm will consist of 200 FLPs, and 250 EPN servers hosting each 8 GPUs and 64 CPU cores. Such computing power was estimated to 
be sufficient to cope with the peak data taking rate of 50 kHz expected during the \PbPb data taking. 

\begin{figure}[ht!]
\centering
\includegraphics[width=1.0\textwidth]{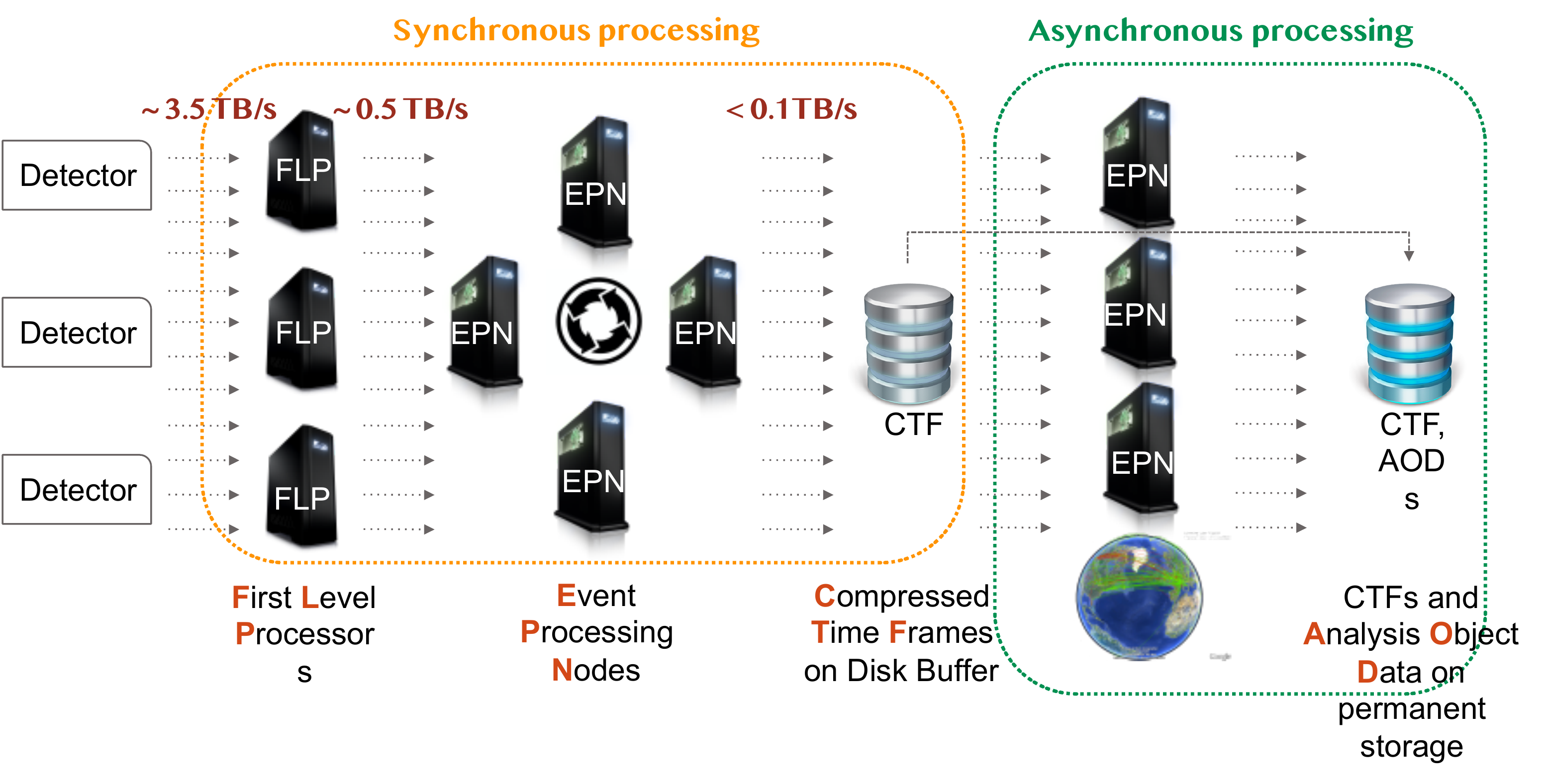}	
\caption{Synchronous and asynchronous processing stages for the ALICE data taking in \run{3} and \run{4}. Data rates refer to the peak interaction
rate foreseen in \PbPb collisions, 50 kHz.}
\label{fig:DataProcessing}
\end{figure}

After a time interval of between 2 and 4 weeks, during which refined calibrations not crucial for data reduction will happen, a second, and eventually third
reconstruction step will take place, in the so called asynchronous processing phase. The reconstruction load will be shared equally between the 
\oo farm and the Tier~0 (T0) and Tier~1 (T1) grid computing nodes. The output of the asynchronous reconstruction will be stored in Analysis Object Data (AODs)
that will be saved to permanent storage and will be the input for physics analysis, while CTFs, previously transferred to the T0 and T1's, will be deleted 
from the disk buffer to make space for new incoming data. Figure~\ref{fig:DataProcessing} shows schematically the ALICE processing stages for \run{3}
and \run{4}.

It is worth to note that the same type of processing will be adopted for both pp and heavy-ion collisions, with the additional implementation of 
software, physics-oriented triggers during the asynchronous reconstruction in the case of pp, to be able to cope with the expected data sizes and rates. 
This means that, for pp data, CTFs will be rewritten to contain only the interesting events at the end of the asynchronous stage, when also AODs will be produced.

\subsection{TPC calibration and reconstruction}
\label{sec: TPC calibration and reconstruction}
As mentioned before, the goal of the synchronous reconstruction will be to reach a factor of 35 in data reduction. The main concerned detector will be 
the TPC, which will have to go down from $\sim$3.4 TB/s to $\sim$70 GB/s. The TPC data compression will consist of different ingredients, among which 
zero suppression, clusterization and optimized cluster data format, entropy reduction and encoding~\cite{MichaelLettrich}, and 
tracking (aimed at removing clusters not associated to signal tracks and further improve the data format used for clusters). 
This challenging goal will be attained through an intense use of GPUs, which 
ensure the necessary computing power at limited costs to process the data synchronously~\cite{MatteoConcas}, and which will also be exploited during
the asynchronous stage.

Despite of the fact that the design and configuration of the TPC GEMs will reduce the ion back flow to the minimum, positive charge will never the less 
accumulate and move
in the TPC drift volume, modifying the electric field, and resulting in space point distortions affecting the TPC clusters. This effect was seen already in \run{1} and 
\run{2}~\cite{ALICE:2014qrd}, even though with different magnitudes and patterns in the $r$ and $z$ directions and $dr\phi$\footnote{The $r$ direction
is the radial one, while $z$ is the longitudinal direction along the beam.} from 
those expected in \run{3} and \run{4}. The main ingredient to the 
calibration strategy will be the same as in \run{2}~\cite{Schmidt:2020pbe}, i.e. the 
ITS-TPC-TRD (Transition Radiation Detector)-TOF (Time Of Flight) track interpolation. Such processing will happen during the synchronous phase on a 
small fraction ($< 1\%$) of the tracks as a preparatory step for the calibration to be applied during 
the final reconstruction, and it will rely on the tracking in the central barrel of ALICE, which will be one of the main challenges in \run{3} and 
\run{4} for the experiment: at the IR of 50 kHz five collisions will overlap in the TPC drift time ($\sim$100~$\mu$s), and clusters will be produced
in continuous readout mode with an absolute time without a well defined associated $z$ coordinate, and will suffer from the aforementioned position-dependent
distortions. The description of the track reconstruction strategy for \run{3} and \run{4} is detailed in~\cite{Rohr:2018cxc}.
While the track-based interpolation between detectors, requiring to collect and process statistics beyond the synchronous timescale, will serve as a correction for the 
average effects of the space point distortions at the asynchronous stage, 
during the synchronous phase, a reference space point distortion map (from previous reference data) will be scaled by the actual luminosity to provide the
needed corrections.
Additionally, fluctuations 
originating from variations in the number of effective instantaneous event and track multiplicity, will not be caught by neither procedure, but  
will need to be corrected for. This will be critical during the asynchronous processing to guarantee the optimal performance of the TPC, with an 
intrinsic resolution of $\sim$200 $\mu$m. To account for fluctuations, 
the history of the measured charge at the readout plane (digital currents) will be used with different granularity for the synchronous and asynchronous
reconstruction stages, as it can be translated into number and position of ions, and, 
in turn, space charge and distortions.

\section{\oo processing model and analysis framework}
No matter what will be the processing step, the same software framework will be used during \run{3} and \run{4}, developed in collaboration 
with the FAIR Software Group at GSI, and inspired by the ALICE \run{1} and \run{2} HLT~\cite{Rohr:2017nrx}. 
The software framework is based on three layers. The first one is the Transport Layer, which defines that the data will be transported as of FairMQ messages
through the different devices building the ALICE software pattern (topology). This guarantees a low level generalization of the 
software architecture which will be flexible and suitable for different network configurations, with the advantage of a shared memory to be used if more devices sit on the 
same node. The second layer is the Data model, and it defines the detailed description of the messages to be passed, which is computer language agnostic, 
and extendible. It allows for multiple data formats and serialization methods, like custom data structure that may be GPU oriented (useful e.g. 
for reconstruction), ROOT objects (for Quality Control, QC), and Apache Arrow (for analysis). Finally, the Data Processing Layer (DPL) serves as a sort of 
translator of the user's computational problem into a low level topology of devices exchanging messages.  

As all the other processing elements (reconstruction, calibration, QC, simulation), the ALICE analysis framework will be based on the \oo processing model. The 
AOD data will be organized in flat tables (using the Apache Arrow layout behind a classic C++ API) to minimize I/O costs, and allow for exploitation of vectorization and parallelization. The analysis core will be expressed in the form of a task, translated by DPL into FairMQ topologies and devices, 
where filters, selections and data manipulation will be possible, and table will
be concatenated depending on the user needs. Finally, the analysis output will be based on ROOT.

Due to the very high number of data that will be collected (100 more collisions to be analyzed with respect to \run{1} and \run{2}), 
new analysis strategies with respect to the present ones will be adopted in \run{3} and \run{4}. Physics-oriented 
skimming campaigns will allow to selectively store collisions and tracks and reduce the size of interesting data. Besides, an Analysis Facility of 20000
cores will be used for fast analysis validation before the full submission on the entire data sample on the Grid. An organized approach like in \run{1}
and \run{2} will be taken, with analyses organized in ``trains'' (grouping together analyses running on the same data samples) within the ALICE physics 
working groups.

\section{Conclusions}
ALICE is getting ready to a major upgrade for the upcoming \run{3} and \run{4}, touching not only the data taking paradigm and detectors, 
but also the data processing at all stages, from reconstruction to analysis. Data will be first processed synchronously, to achieve an impressive size reduction
factor of 35, allowing to store the data on a disk buffer for the physics-oriented reconstruction stage. Online reconstruction and calibration will play a key role, and the TPC will be the crucial detector. 

The \oo processing
framework based on three layers (data transport, data model, and data processing) will be used in all processing components, from reconstruction, to 
calibration, to analysis. The analysis framework for \run{3} and \run{4} is being built around the computing and storage needs imposed by the very 
large data samples that will be collected, ready to open the door to a new, beautiful physics.

\end{document}